\documentclass{ptapap}

\author{P\i nar Cerraho\u{g}lu}[UD]
\author{V\'eronique Petit}[UD]
\author{Zsolt Keszthelyi}[UA]
\author{Alexandre David-Uraz}[UD]

\affil[UD]{Dept. of Physics and Astronomy, University of Delaware, Newark, DE, USA}
\affil[UA]{Anton Pannekoek Institute for Astronomy, University of Amsterdam, The Netherlands}

\title{The evolution of magnetic stars in a single-age population}

\begin{document}

\maketitle

\begin{abstract}

Observational and theoretical work has now established that the fossil fields of magnetic massive stars are surviving remnants from an earlier event, or an earlier evolutionary phase. However, many important questions remain regarding the effects of these fields on the late stages of stellar evolution, as well as their impact on the core-collapse mechanism and the formation of exotic compact objects such as magnetars and gravitational wave progenitors. There is currently a critical need to incorporate the impact of fossil fields in models of the structure and evolution of magnetic stars, and to determine the evolutionary history of magnetic massive stars. We present a preliminary population study of a cluster of co-evolving stars based on MESA evolutionary tracks that account for the effect of magnetic mass-loss quenching.

\end{abstract}

\section{MESA Models}

The presence of a fossil field in a massive star can alter its structure in significant ways, for example:
(i) Quenching of the mass-loss because of wind material trapped in the closed magnetic loops \citep[e.g.][]{2002ApJ...576..413U, 2017MNRAS.466.1052P, 2017A&A...599L...5G}.
(ii) Loss of angular momentum due to magnetic braking \citep[e.g.][]{2009MNRAS.392.1022U, 2011A&A...525L..11M, 2019MNRAS.485.5843K}
(iii) Suppression of internal convection and/or core overshoot \citep[e.g.][]{2012MNRAS.427..483B, 2013MNRAS.433.2497S, 2019MNRAS.487.3904M}

Here, we concentrate only on the evolutionary impact of the fossil field on the mass-loss due to magnetic confinement. We thus use non-rotating models that include mass-loss quenching under the assumption that the magnetic flux is conserved with time.

Our models have been calculated with the Modules for Experiments in Stellar Astrophysics \citep[MESA;][]{2011ApJS..192....3P,2013ApJS..208....4P,2015ApJS..220...15P}
 using the non-rotating prescriptions described in \citet{2017MNRAS.466.1052P} and \citet{2019MNRAS.485.5843K}. 
Our grid ranges from 5\,M$_\odot$ to 63\,M$_\odot$ in increments of 0.1 dex, and from 630\,G to 15.8\,kG in 0.2 dex increments. A model with $B_p=0$ is also included. 

\section{Population Study}

Our goal is to simulate a single burst of star formation, and study how the resulting population characteristics change with cluster age. These characteristics, under various assumptions about the birth properties of magnetic stars, can then be compared to the properties of known magnetic massive stars. For this initial study, we investigate whether various “Initial magnetic B-field Functions” \cite[IBFs,][]{2019MNRAS.489.5669P} can reproduce the general $\sim 10\%$ incidence of detected magnetic stars \citep{2016MNRAS.456....2W, 2019MNRAS.483.3127S}.

We create a cluster of massive stars that contains 200 stars. We use a fixed Initial Mass Function (IMF) following a Salpeter \citep{1955ApJ...121..161S} power-law with an index of -2.35. This results in 2 stars in the most massive bin, and 55 stars in the least massive bin. 
For each realization of the cluster, we randomly assign an initial magnetic field value to each star in the cluster, following (i) a linearly uniform distribution of field strengths between 0 and 30 kG, representing a lack of knowledge about the IBF\footnote{An alternative sampling distribution could assign the same probability per decade of field strength.}, and (ii) a power-law distribution using two arbitrary values for the index ($-0.8$ and $-1.2$), illustrative of a specific type of IBF that could be tested. We note that in order to normalize the probability function that generates the power-law IBFs, we bound our range of magnetic strength values between 10 G and 30 kG, such that $\int_{B_{p,\mathrm{min}}}^{B_{p,\mathrm{max}}}P(B_p) dB_p=1$. As our grid of evolution models is quantized, we use the model for the closest lower field strength value, and the $B_p=0$ model is used the the lowest field strength bin.

As a first approximation, we conservatively assume that a surface magnetic field with a dipolar strength of 500 G can be detected at any time \citep[slightly larger than the typical upper limit of the O-type star survey of the MiMeS project;][]{2019MNRAS.489.5669P}. More detailed detection limits and observational biases will be included in this type of study. We also assume that the hypothetical survey observing these clusters only includes stars on the main sequence.

As a function of cluster age, we compute the number of magnetic stars that would be detected in the cluster. As a reminder, our evolution models include the evolutionary feedback of the mass-loss quenching on the radius of the star and hence on the surface magnetic field strength as a function of time.

\begin{figure}
\includegraphics[width=\textwidth]{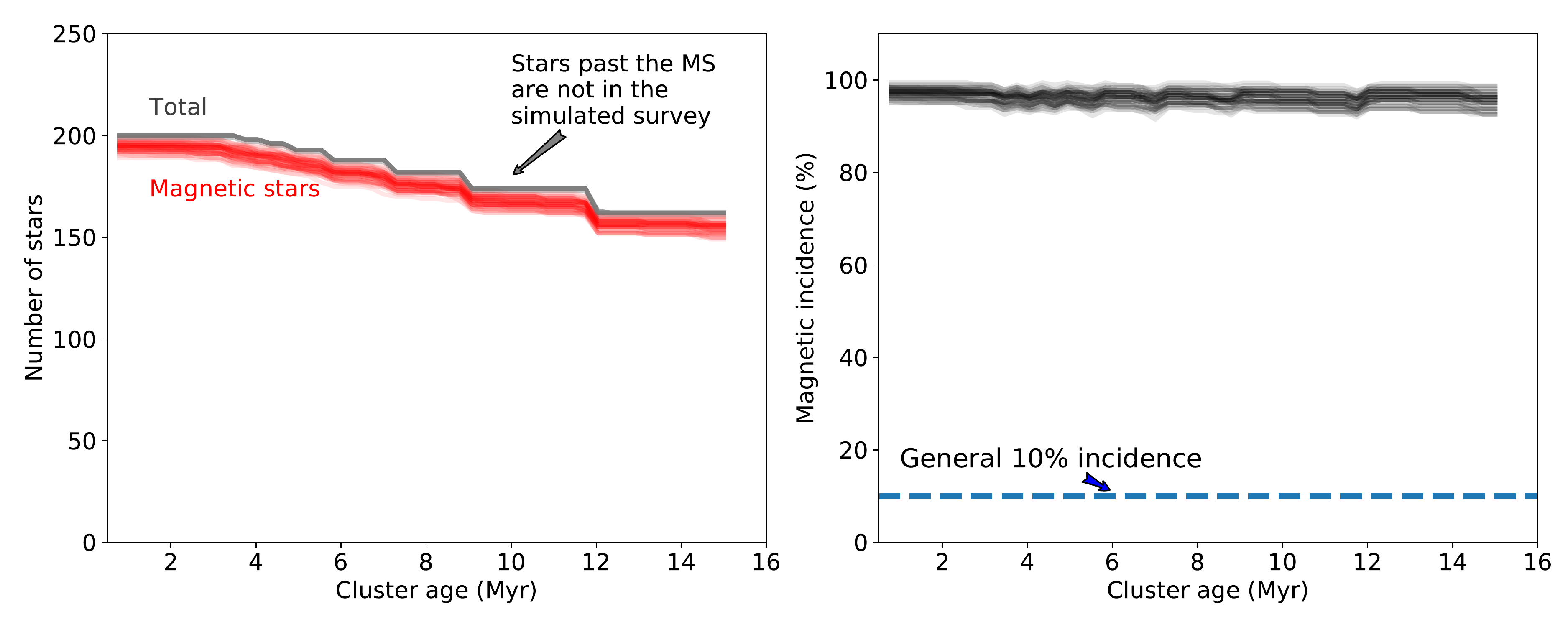}\\
\includegraphics[width=\textwidth]{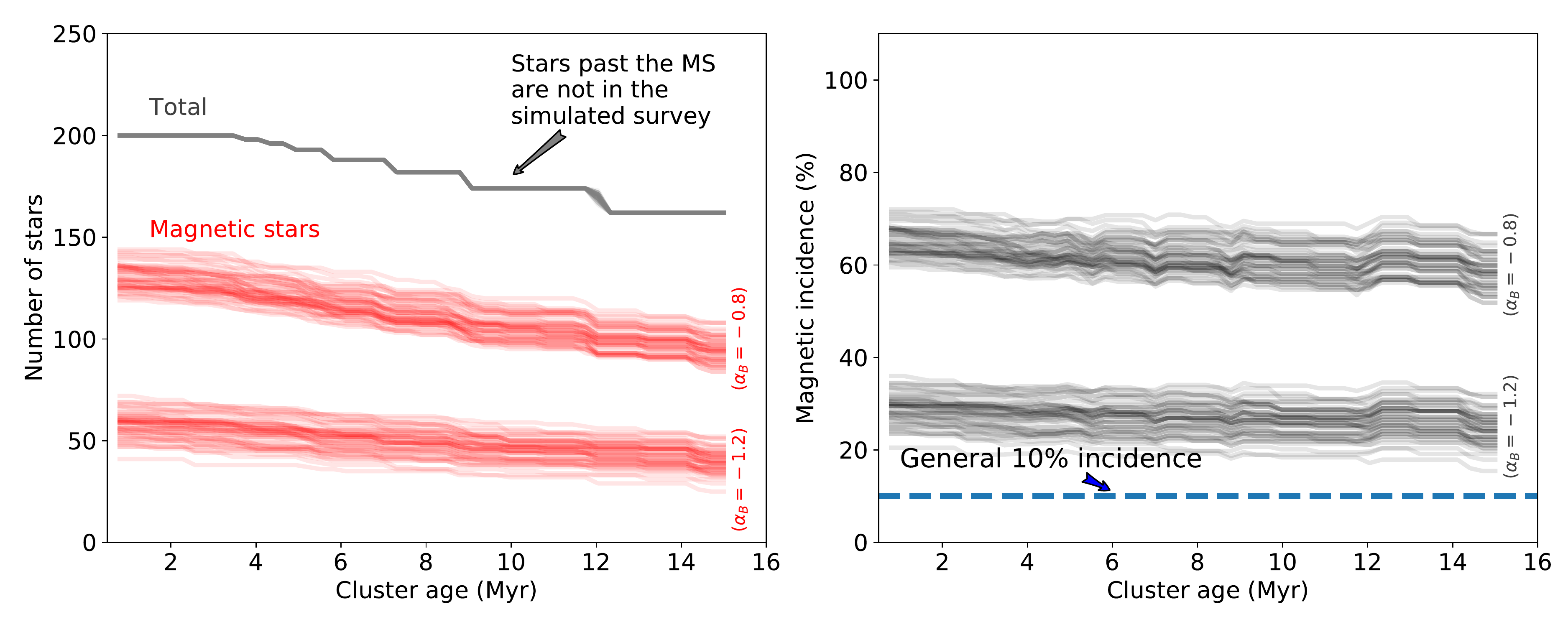}
\caption{\label{fig:fig1} Total number of stars that are still on the main sequence in the simulated cluster (left-hand panels, grey) total number of stars with a detectable magnetic field (i.e. $B_p > 500$\,G) (left-hand panels, red) and incidence of magnetic stars (right-hand panels) as a function of cluster age. The observed general 10\% incidence is indicated with a blue dashed line in the right panels. Each curve corresponding to a single simulation has a transparency of 10\%, so that darker colors represent regions of overlap. The top panels show the case of a uniform IBF and the bottom panels show the case of a power-law IBF with indices $\alpha_B$ of $-0.8$ and $-1.2$.}
\end{figure}

\section{Results}

For each IBF distribution, we simulate 50 clusters. In Fig. \ref{fig:fig1}, each resulting curve is transparent, thus darker color represents regions of overlap. 

The top row of Fig.\,\ref{fig:fig1} presents the case of a uniform IBF distribution. In the left panel, the grey curves show the total number of stars on the MS in the cluster as a function of cluster age, up to 15 Myr (which corresponds to the time at which a 12\,M$_\odot$ star leaves the MS in our models). The number of stars decreases because on stars on the MS are considered in our simulated surveys. The red curves show the number of detectable magnetic massive stars. In the right-hand panel, we show that the incidence of massive stars with a detectable magnetic field is nearly constant with stellar age, but much larger than the 10\% general incidence of magnetism in OBA stars illustrated by the dashed blue horizontal line. Thus such a uniform IBF distribution is clearly not compatible with the general properties of magnetic massive stars. 

The bottom row of Fig.\,\ref{fig:fig1} presents the cases of two power-law distributions for the IBF, with indices of $-0.8$ and $-1.2$. The lower (more negative) index value results in a general incidence that is closer to the observed 10\% incidence, when compared to a higher power-law index and to the uniform IBF distribution. 

For IBF distributions that reproduces the overall incidence, we can study other population characteristics, such as the independence of the incidence with respect to mass, as well as the distribution of current-day field strengths which should peak at a few kilogauss \citep{2019MNRAS.490..274S}. Finding suitable IBFs that reproduce all of these properties simultaneously will put strong constraints on formation scenarios and help recreate the full evolutionary history of these stars, from birth to death.

\acknowledgements{VP and PC acknowledge support from the University of Delaware Research Foundation (UDRF). ADU acknowledges support from the Natural Sciences and Engineering Research Council of Canada (NSERC).}

\bibliographystyle{ptapap}
\bibliography{database}

\end{document}